\documentclass[aps,pra,showpacs,showkeys,onecolumn,superscriptaddress, notitlepage]{revtex4-1}

\usepackage{graphicx,tabularx,times,hyperref}
\usepackage{caption}
\setcitestyle{numbers,square}
\usepackage{float}

\usepackage{array}
\usepackage{amsmath}
\usepackage{units}
\usepackage{graphicx}
\usepackage{braket}
\usepackage{xfrac}
\usepackage{enumitem}
\usepackage[normalem]{ulem}
\usepackage{booktabs}
\usepackage{tabularx}
\usepackage{multirow, bigstrut}
\usepackage{hhline}
\usepackage{times}

\begin{document}

\title{Comparing introductory physics and astronomy students' attitudes and approaches to problem solving}
\author{Melanie Good}
\affiliation{Department of Physics and Astronomy, University of Pittsburgh, Pittsburgh, PA 15260}
\author{Andrew Mason}
\affiliation{Department of Physics and Astronomy, University of Central Arkansas, Conway, AR 72035}
\author{Chandralekha Singh}
\affiliation{Department of Physics and Astronomy, University of Pittsburgh, Pittsburgh, PA 15260}

\begin{abstract}
Students taking introductory physics and introductory astronomy classes, which are both gateways to a physics or physics and astronomy major, may have different attitudes and approaches to problem solving.  We examined how introductory physics students' attitudes and approaches to problem solving compare to those of introductory astronomy students, using  a previously validated survey, the Attitudes and Approaches to Problem Solving (AAPS) survey. In addition, we compared the performance of the introductory physics and astronomy students on the factors which were identified in a factor analysis in the original validation study. 
We found that introductory astronomy students' overall average AAPS score was significantly more favorable than that of introductory physics students ($p<0.01$), and the effect size was large (Cohen's d = 0.81).  We also found that introductory astronomy students' scores were more favorable in all clusters of questions except for one factor involving drawing diagrams and writing scratchwork while solving problems.  Follow-up interviews suggest that one possible explanation for less favorable scores in this factor is the context of astronomy problems, e.g., the difficulty and usefulness of drawing electromagnetic radiation. Moreover, introductory astonomy students who were interviewed indicated that they would likely draw diagrams for problems that lend themselves well to sketching, such as problems involving celestial mechanics. We also found that introductory physics and astronomy students were equally capable of solving two isomorphic problems posed to them, and that the majority of introductory physics and introductory astronomy students reported that the problem posed in the astronomy context was more interesting to them. Interviews suggest that the context of astronomy in problem solving may be more interesting for students and could be one possible explanation for the more favorable AAPS scores amongst introductory astronomy students compared to introductory physics students.  Instructors of introductory physics courses should heed these findings which  indicate that it may be beneficial for instructors of introductory physics courses to incorporate problems into their instruction which contain real-world contexts, which may serve to increase student interest-level, and which could help create more favorable attitudes and approaches towards problem solving.
\end{abstract}

\maketitle

\section{Introduction}

\subsection{Differences between introductory physics and introductory astronomy classes and students}

A typical calculus-based or algebra-based introductory physics course may offer a different experience for students compared to a typical introductory astronomy course.  
 Although both courses are intended for science and engineering majors and they cover many physics principles underlying the relevant content that are the same, the content of the two courses is generally organized around very different themes.  For example, the introductory physics course often focuses on a linear progression in terms of introducing students to various concepts and fundamental laws one by one, often in a bottom-up approach. In particular, in a typical algebra-based or calculus-based physics course for biological science, physical science, and engineering majors, students are typically introduced to vectors and various kinematic and dynamic variables.  This is generally followed by coverage of Newton's laws of motion, impulse and momentum, and work and energy.  Then students learn about rotational kinematics, dynamics, and angular momentum followed by simple harmonic motion, gravitation and waves.  The treatment of these topics includes some calculus for the calculus-based students who are physical science and engineering majors compared with the algebra-based students who are typically biological science majors. Meanwhile in a typical introductory astronomy course for science and engineering majors discussed here (which is mandatory only for ``physics and astronomy" majors), students learn about observational techniques, stars and stellar evolution, and the interstellar medium.  This is followed by study of galaxies and cosmology. In a typical introductory astronomy course for science and engineering majors similar to the one we focus on here, the material is taught in a quantitative fashion with astrophysical problems requiring application of underlying physical principles.  While there are some topics that are unique to this introductory astronomy (e.g., properties of light), there is significant overlap of the physics principles utilized in an introductory astronomy course compared with an introductory physics course.  While studying the astronomy topics listed above, students learn physics principles such as Newton's laws of motion, conservation laws (e.g., conservation of energy, conservation of angular momentum etc.), rotational motion, and gravitation.  

While both courses are recommended for science and engineering majors, the courses are often taught differently. From the manner in which this type of introductory astronomy course for science and engineering majors is organized, it is clear that the focus is on understanding the large-scale structure of the universe and the rules that govern the past and future evolution of astrophysical objects.  In an introductory physics course, the focus is often on helping students learn to apply the laws of physics in simplified contexts, e.g., applying Newton's laws to blocks on inclined planes or masses connected via ropes and pulleys. Partly due to the focus of the course, astronomy classes often involve the use of many photographic images of objects in space; whereas physics classes often only rely upon sketches, cartoons, or other abstract representations of the objects being studied.

\subsection{The role of expertise in problem solving}

Experts, such as physics faculty members, organize their knowledge hierarchically, such that underlying concepts which are related are connected in a meaningful and structured way \cite{Chi, Reif, DeMul, partial, Harper, Gladding}.  This knowledge structure allows experts to efficiently approach the problem-solving process \cite{Reif}.  By contrast, novices, who are trying to develop expertise, such as introductory students, may view physics as a collection of disconnected facts and equations \cite{Chi, Reif}.  The lack of organization to their knowledge structure can result in introductory students approaching each problem as a unique challenge in which they search for the correct formula or resort to ``plug and chug'' \cite{Reif}. Within the spectrum of expertise devlopment, students can span various levels of expertise \cite{MasonCat}, and their level of expertise may be connected to their attitude and approach to solving problems. Moreover, the approaches and attitudes students use in their learning and problem-solving can impact the extent to which they take the time to organize their knowledge structure \cite{Schommer, Elby3, Cummings, Redish, Chi, Reif}.  This can, in turn, influence the acquistion of conceptual understanding and problem-solving skills \cite{Schommer, Elby3, Cummings, Redish}. Thus examining where on the continuum of expertise students lie \cite{MasonCat} can serve to illuminate the extent to which they have developed conceptual understanding and problem-solving skills, and measuring their attitudes and approaches to problem-solving may be one way to examine their place on the continuum of expertise.

\subsection{Assessing introductory physics students' views using attitudinal surveys}

Understanding students' attitudes and approaches to problem solving is important because it can have instructional implications. 
Attitudinal surveys have been developed to assess students' beliefs about physics.  The Maryland Physics Expectation Survey (MPEX) and the Colorado Attitudes about Science Survey (CLASS) are similar in that they focus on assessing attitudes students have about physics \cite{Redish, Adams}. The Epistemological Beliefs Assessment for Physics Science (EBAPS) survey was designed and validated to probe purely epistimological stances of students of physical sciences along multiple dimensions \cite{Elby}. The Attitudes toward Problem Solving Survey (APSS) was developed through inspiration from the MPEX survey, with a focus on attitudes about problem solving \cite{Marx, Cummings}.  

A modified version of the APSS, the Attitudes and Approaches to Problem Solving (AAPS) survey, was developed to include questions regarding the approaches students take when solving problems \cite{Mason1}.  This survey was validated based upon faculty, graduate student, and introductory student responses \cite{Mason1}.  
During validation, expected ``favorable'' responses agreed with faculty responses to a high degree, and it was found to be the case that, on average, introductory students responded differently from graduate students and faculty, in that introductory student responses were found to be less ``favorable'' compared to graduate students and faculty \cite{Mason1}.

\subsection{Focus of our research}

The focus of our research presented here was on analyzing, comparing, and interpreting the Attitudes and Approaches to Problem Solving (AAPS) survey responses for introductory algebra-based and calculus-based physics and introductory astronomy students. The survey was administered  as a posttest at the end of the course. 
All the courses involved in this investigation (whether physics or astronomy) were intended for science and engineering majors, which means that students from these majors constitute the overwhelming majority of students enrolled. Our research questions are as follows:

RQ1. Are there differences in average overall scores on the AAPS survey for introductory astronomy students compared with introductory physics students?

RQ2. Are there differences in scores on specific clusters of questions on the AAPS survey for introductory astronomy students compared with introductory physics students?

RQ3. When presented with an isomorphic problem pair (problems with the same underlying physics principle but different contexts--one written in an astronomy context and the other written in a physics context), are introductory physics and introductory astronomy students equally likely to solve both problems correctly and do they find either more or less interesting?

\section{Methodology}

\subsection{Courses and participants}

The courses in our study consisted of both algebra-based and calculus-based introductory physics, and introductory astronomy at the University of Pittsburgh.  Introductory physics is often mandatory for most science and engineering majors (who constitute a majority of students in the class). The introductory astronomy course for science and engineering majors is taken by many students as an elective although it is mandatory for ``physics and astronomy" majors. We note that a majority of students in the astronomy course discussed here are science or engineering majors since it is listed in course listing as a course for science and engineering majors and others are advised to enroll in another course with similar content but lower mathematical rigor.  
Also, since all physical science and engineering majors at the University of Pittsburgh are required to take a two-course sequence in introductory calculus-based physics in their freshman year, some students in the astronomy class may have already taken introductory physics. Note that no statistically significant difference was found in the AAPS scores for algebra-based introductory physics (taken mainly by biological sciences majors) compared with calculus-based introductory physics, so these two types of classes were combined. 
We note that the calculus-based physics classes are required for engineering majors but physics majors have the option to take an honors version of these classes if they wish, which were not included in this study. Thus, not all physics majors take the classes included in this study. A total of 606 students participated in this study.  This includes 541 introductory physics students (including algebra-based and calculus-based courses) and 65 introductory astronomy students.  
The AAPS survey responses from graduate students and faculty are included in order to serve as a benchmark of more expert-like responses \cite{Mason1}. A total of 42 graduate students and 12 faculty responses to the AAPS survey are included. A subset of the total sample of students participated in individual follow-up interviews, each lasting approximately 1 hour.  This included 12 introductory physics students and 8 introductory astronomy students.

\subsection{Data collection}

The AAPS survey was given to students in this study near the end of the semester.  The items in the AAPS survey are statements which can elicit agreement or disagreement, and responses are given on a 5 point Likert scale (strongly agree, agree, neutral, disagree, or strongly disagree). When developed and validated, the items were designed so that a favorable response is not always ``agree'' or ``disagree.'' Once completed, spreadsheets of the results were produced and scores for each question were computed by assigning a +1 to a favorable response, a -1 to an unfavorable response, and a 0 to a neutral response, in which ``favorable'' means that the response reflects that which an expert in the field (such as a faculty member) would give. We will refer to this convention (of scoring between -1 and +1) as ``normalized'' data, which we adopted for consistency with the convention used when the AAPS survey was originally validated \cite{Mason1}.  We then averaged these values across each group of interest (e.g., introductory physics or introductory astronomy). This average score will be referred to as the ``normalized'' score.  In addition, we also computed the percentage of the total responses that were favorable, neutral, and unfavorable responses for each question, and we also averaged these over all questions to obtain the average scores for each question on the AAPS survey.  These percentages (of favorable, unfavorable, and neutral responses) will be referred to as ``unnormalized'' scores. The average unnormalized favorable, unfavorable, or neutral scores on a particular question refer to the percentage of students who had favorable, unfavorable, or neutral responses on that question.  The average unnormalized favorable, unfavorable, or neutral scores for the entire AAPS survey refer to the average of the unnormalized scores (of favorable, unfavorable, and neutral responses) for all questions on the AAPS survey.   

\begin{figure}[!htbp]
\includegraphics[width=1\textwidth,trim = {0 12.0cm 0 12.0cm}]{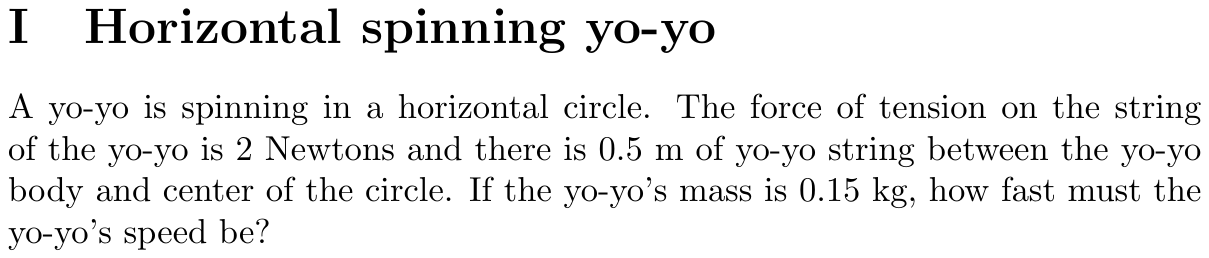}
\caption[justification=raggedright]{Non-astronomy context in isomorphic problem pair. Note that students were told to assume no other forces were present other than the force of tension.}
\label{fig:IPP1}
\end{figure}

\begin{figure}[!htbp]
\includegraphics[width=1\textwidth,trim = {0 12.0cm 0 12.0cm}]{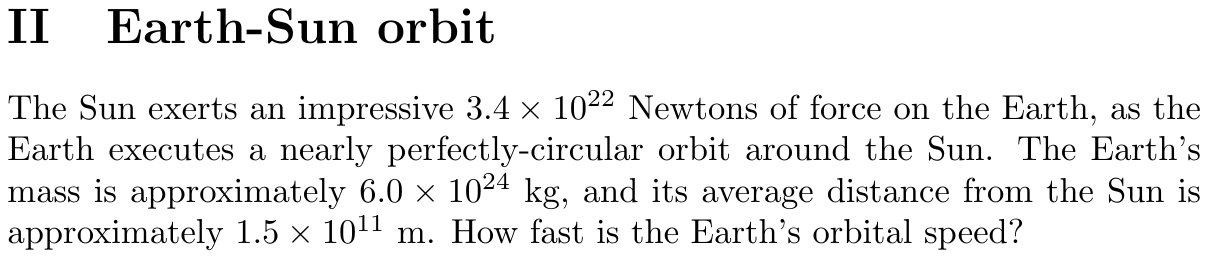}
\caption{Astronomy context in isomorphic problem pair.}
\label{fig:IPP2}
\end{figure}

In order to gather qualitative data, the written survey data collection was followed by individual hour-long interviews with some introductory physics and introductory astronomy students.  The interviews utilized a ``think aloud'' protocol in which students answered the AAPS survey questions along with providing reasoning for why they answered the way they did.  They were not disturbed as they answered the survey questions while thinking aloud, but later we asked them for clarifications of the points that had not been made clear. Follow up questions were asked when necessary in order to probe deeper into students' reasoning.  In addition, in the interviews, both introductory physics and introductory astronomy students were presented with an isomorphic pair of problems. As can be seen in Figures \ref{fig:IPP1} and \ref{fig:IPP2}, one was written in a non-astronomy context and the other written in the context of astronomy.  They were asked to solve both problems and were asked about whether either problem was more difficult or enjoyable.  Both problems required the students to solve for the speed of an object based upon the assumption of uniform circular motion with centripetal acceleration.  Thus, the problems required the same concepts and formulas, but contained different contexts--the astronomy problem involved the motion of the Earth around the Sun and the physics problem involved the circular motion of a yo-yo whirled in a horizontal circle.  The centripetal acceleration in each case was due to different mechanisms (gravity for the astronomy problem and tension for the yo-yo problem), but the manner of solving the problems is essentially identical.  The purpose was to determine if both physics and astronomy students were equally proficient in solving these problems, whether one of the problems was more challenging than the other and whether there was something about the context of astronomy or physics that produced different attitudes or approaches to the problem-solving process. We also looked at the pre-/post- test data for the Force Concept Inventory (FCI), a standardized conceptual survey \cite{Savinainen}, to investigate how the performance of physics and astronomy students compares on this topic, which is taught in both classes.

\section{Results}

\subsection{RQ1. Are there differences in average overall performance on the AAPS survey for introductory astronomy students compared with introductory physics students?}

\begin{figure}[!htbp]
\includegraphics[width=0.60\textwidth]{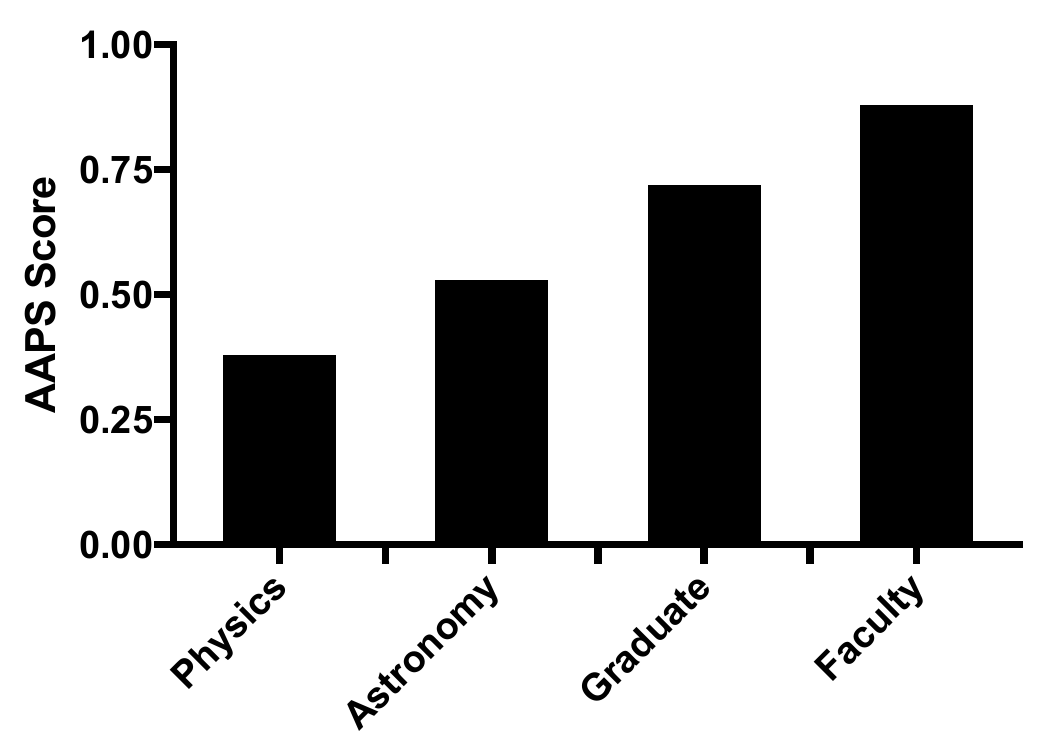}
\caption[justification=raggedright]{Average AAPS scores of introductory physics students, introductory astronomy students, graduate students, and faculty.}
\label{fig:Figure1}
\end{figure}

 Figure \ref{fig:Figure1} shows that introductory physics and astronomy students both have less expert-like attitudes and approaches to problem solving, compared with graduate students and faculty. However, astronomy students have more expert-like attitudes and approaches than physics students. The comparison to graduate students and faculty is based upon the results from the validation study \cite{Mason1}.   
All differences in overall scores were found to be statistically significant on the t-tests, with $p<0.001$ in all cases.  The effect size, as defined by Cohen's d, is given by: $d = \frac{\mu_1 - \mu_2}{\sigma_{pooled}}$ where $\mu_1 - \mu_2$ is the difference in means between two groups and $\sigma_{pooled}$ is the pooled standard deviation, given by $\sigma_{pooled} = \sqrt{\frac{(n_1-1)\sigma_1^2+(n_2-1)\sigma_2^2}{n_1+n_2-2}}$ where $n_1, n_2, \sigma_1$, and $\sigma_2$ are the sample sizes and standard deviations of the two groups being compared \cite{Cohen} in Table \ref{table:Table0}.

As noted, while the content is presented differently in introductory physics and astronomy classes, many of the underlying physics principles involved in the instruction are the same.  In particular, while the courses are presented very differently, foundational topics in physics, such as forces, energy and angular momentum etc. are part of the content of both courses. Indeed, when we compare the performance on the Force Concept Inventory (FCI), we find that students in introductory astronomy had very similar pre- and post-test scores (taken at the beginning and end of the semester, respectively) compared to introductory physics students at the same institution, for a typical cohort of introductory physics and astronomy students included in this investigation.  In particular, introductory physics students scored an average of 47\% on the FCI pre-test, and an average of 58\% on the FCI post-test.  Introductory astronomy students scored an average of 46\% on the FCI pre-test, and an average of 62\% on the FCI post-test.  Thus, the scores on force concepts 
is comparable.  Therefore, the difference in attitudes and approaches to problem-solving in physics and astronomy is not likely to be due to the differences in treatment of concepts such as Newtonian mechanics but some other reason.  For example, there may be something about the manner and context in which introductory astronomy classes are taught which might at least partly account for this difference. In other words, while introductory astronomy students learned just as many physics concepts pertaining to forces and Newton's laws as introductory physics students as measured by the FCI, introductory astronomy students appear to have more favorable attitudes than introductory physics students, perhaps owing to a more engaging context in which the content is learned and the manner in which the entire course curriculum is framed. 

Some evidence for the lack of an engaging context was seen in individual student interviews with physics and astronomy students.  For example, one student expressed that he would enjoy physics classes more if he could relate them to something interesting or realistic.  He went on to point out a common type of introductory physics problem as a counter-example to an interesting or realistic type of problem, stating ``How many times are you realistically doing that with a pulley?''  Another student, in an interview, discussed at length how she had the opportunity to study modern physics topics in high school, and lamented that introductory physics was not presented in an interesting way, stating: ``I feel like there's definitely more about physics, and that this is just the rut of it... before you can get to more interesting things.''  Although this student appeared to assume that the topics would likely get more interesting once you progress beyond introductory physics (based upon the modern physics she learned in high school), it was clear from the discussion that, as it stands, introductory physics was not engaging or interesting in her opinion.

Similarly, a student who had taken both introductory physics and introductory astronomy expressed excitement when comparing the two experiences, stating: ``Astronomy is more exciting.  I just read this story today about a galaxy that doesn't have any dark matter.  That's freaking awesome... Physics doesn't do that.  Anyone can learn about Newton's laws of motion.''  Further discussions suggested that not only did the astronomy course generate his interest in astronomy, but he took the initiative to read news stories related to astronomy on his own. He also hinted at the contrasting way in which introductory physics is taught, in which new discoveries and exciting events were not part of typical instruction.

\subsection{RQ2. Are there differences in performance on specific clusters of questions on the AAPS survey for introductory astronomy students compared with introductory physics students?}

There are a few AAPS survey questions in which physics students outperformed astronomy students. However, for most of the AAPS survey questions, astronomy students outperformed physics students.  This can be seen in Table \ref{table:Table1}, which contains normalized scores on each question, and in Table \ref{table:astrophyrawQ} as raw percentages of favorable, neutral, and unfavorable responses. A factor analysis was performed earlier in the AAPS validation study \cite{Mason1}.  Table \ref{table:Table2} compares the factors identified in the validation study \cite{Mason1} with responses from introductory physics and astronomy students.  Astronomy students generally outperformed physics students in all factors except for Factor 2, which is related to drawings and scratchwork while problem solving.  

\begin{figure}
[!htbp]
\includegraphics[width=0.60\textwidth]{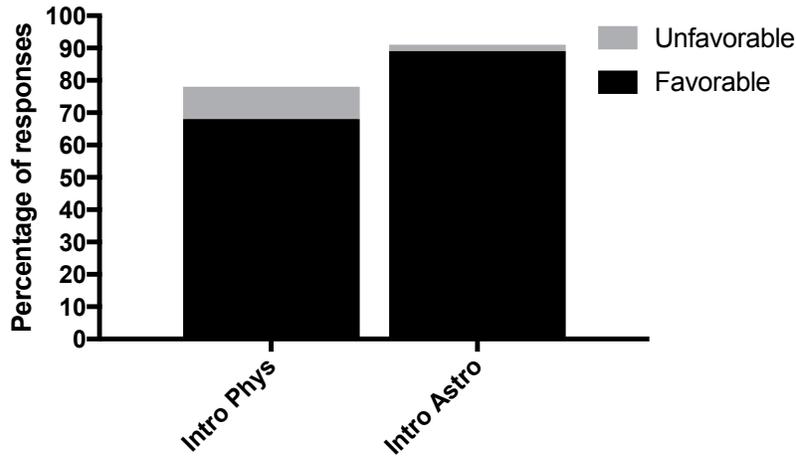}
\caption[justification=raggedright]{Unnormalized scores for Question 10: Astronomy students exhibited more favorable attitudes than physics students when asked if they reflect on principles when they are stuck.}
\label{fig:Q10}
\end{figure}

\begin{figure}
[!htbp]
\includegraphics[width=0.60\textwidth]{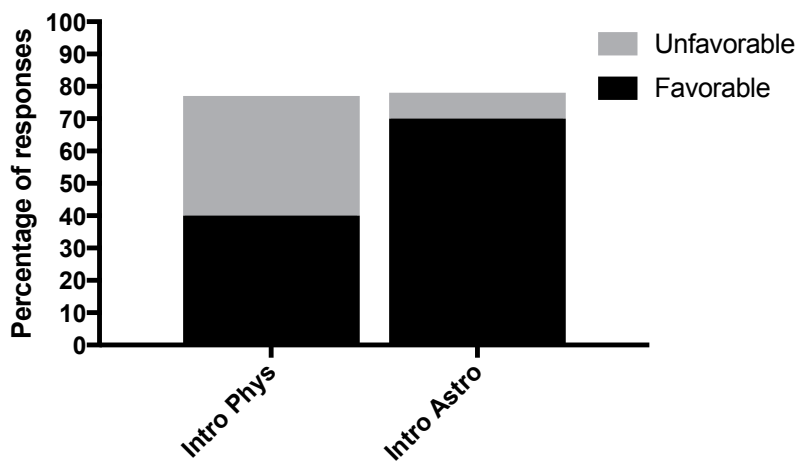}
\caption[justification=raggedright]{Unnormalized scores for Question 27: Astronomy students exhibited more favorable attitudes than physics students when asked if they enjoy solving problems.}
\label{fig:Q27}
\end{figure}

\begin{figure}
[!htbp]
\includegraphics[width=0.60\textwidth]{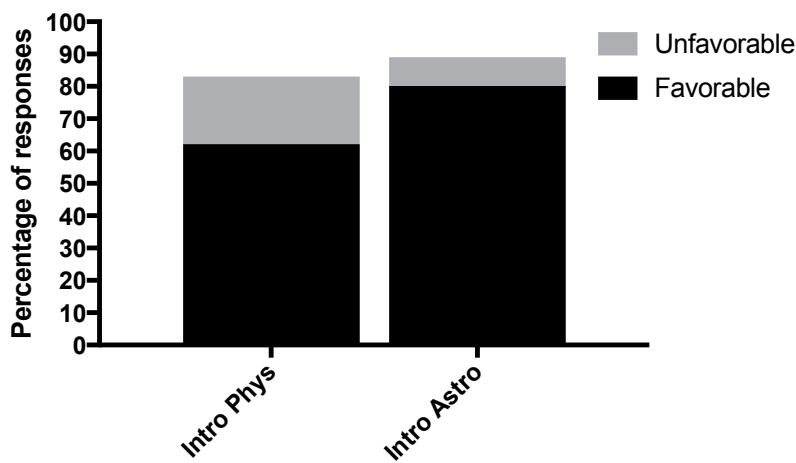}
\caption[justification=raggedright]{Unnormalized scores for Question 4: Astronomy students exhibited more favorable attitudes than physics students when asked if they identify principles before looking for equations.}
\label{fig:Q4}
\end{figure}

\begin{figure}
[!htbp]
\includegraphics[width=0.60\textwidth]{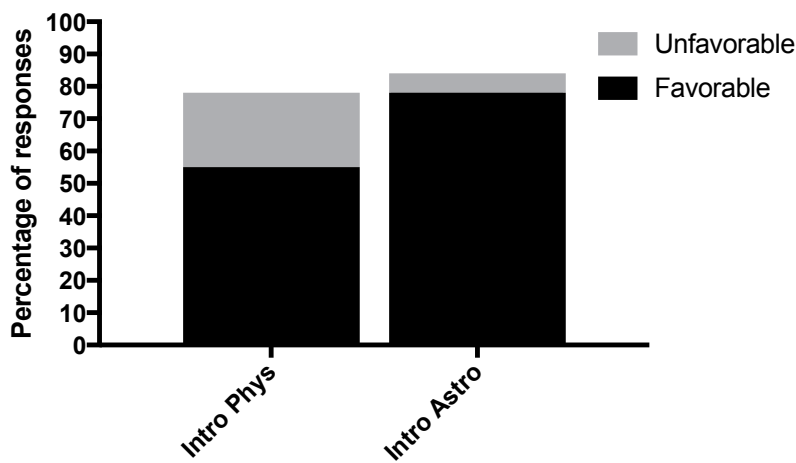}
\caption[justification=raggedright]{Unnormalized scores for Question 14: Astronomy students exhibited more favorable attitudes than physics students when asked if they think about the concepts underlying the problem.}
\label{fig:Q14}
\end{figure}

\begin{figure}
[!htbp]
\includegraphics[width=0.60\textwidth]{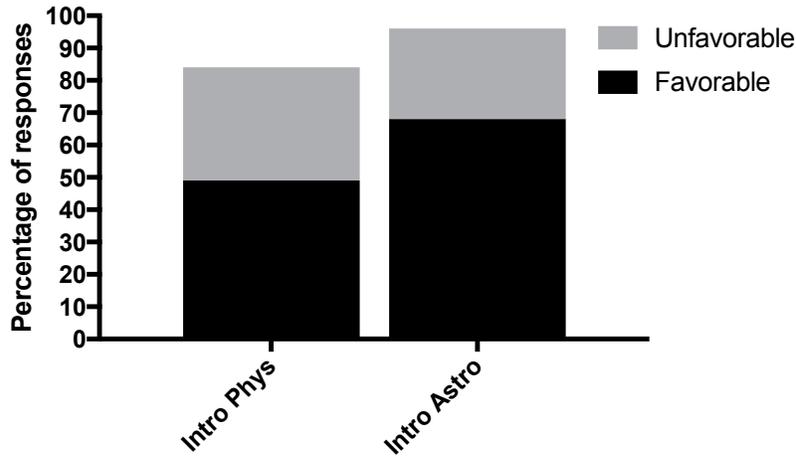}
\caption[justification=raggedright]{Unnormalized scores for Question 1: Astronomy students exhibited more favorable attitudes than physics students when asked if they are stuck unless they go see the teacher or TA.}
\label{fig:Q1}
\end{figure}

\begin{figure}
[!htbp]
\includegraphics[width=0.60\textwidth]{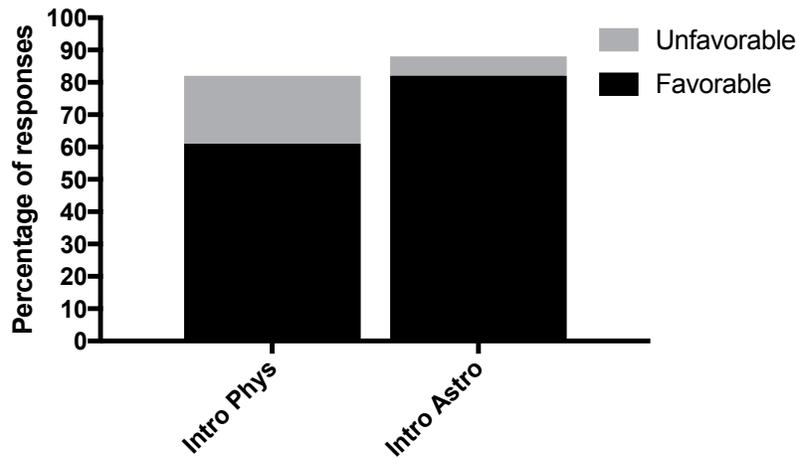}
\caption[justification=raggedright]{Unnormalized scores for Question 23: Astronomy students exhibited more favorable attitudes than physics students when asked if they give up on a problem after 10 minutes.}
\label{fig:Q23}
\end{figure}

\begin{figure}
[!htbp]
\includegraphics[width=0.60\textwidth]{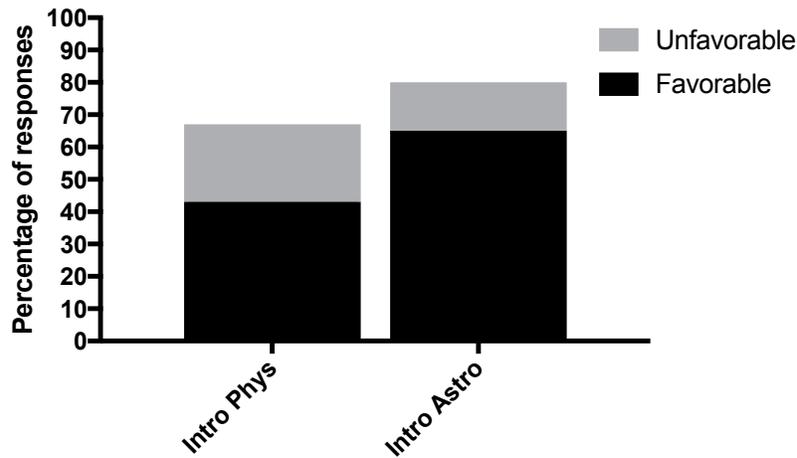}
\caption[justification=raggedright]{Unnormalized scores for Question 2: Astronomy students exhibited more favorable attitudes than physics students when asked if they often make approximations.}
\label{fig:Q2}
\end{figure}

\begin{figure}
[!htbp]
\includegraphics[width=0.60\textwidth]{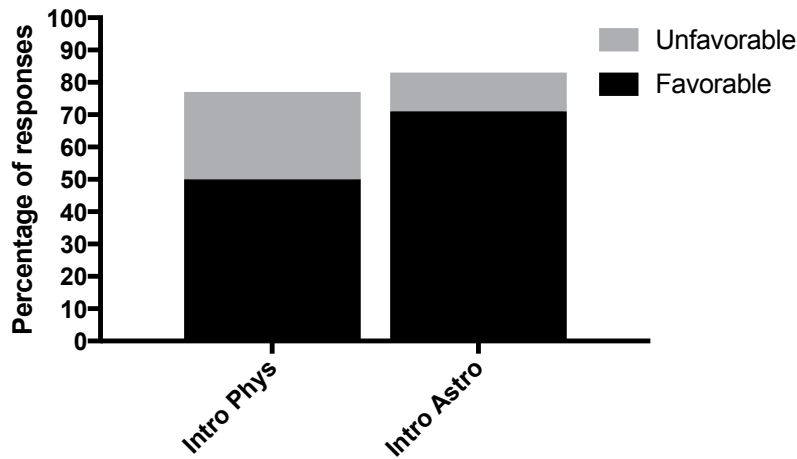}
\caption[justification=raggedright]{Unnormalized scores for Question 16: Astronomy students exhibited more favorable attitudes than physics students when asked if they mostly use their gut feeling when answering conceptual questions.}
\label{fig:Q16}
\end{figure}

\begin{figure}
[!htbp]
\includegraphics[width=0.60\textwidth]{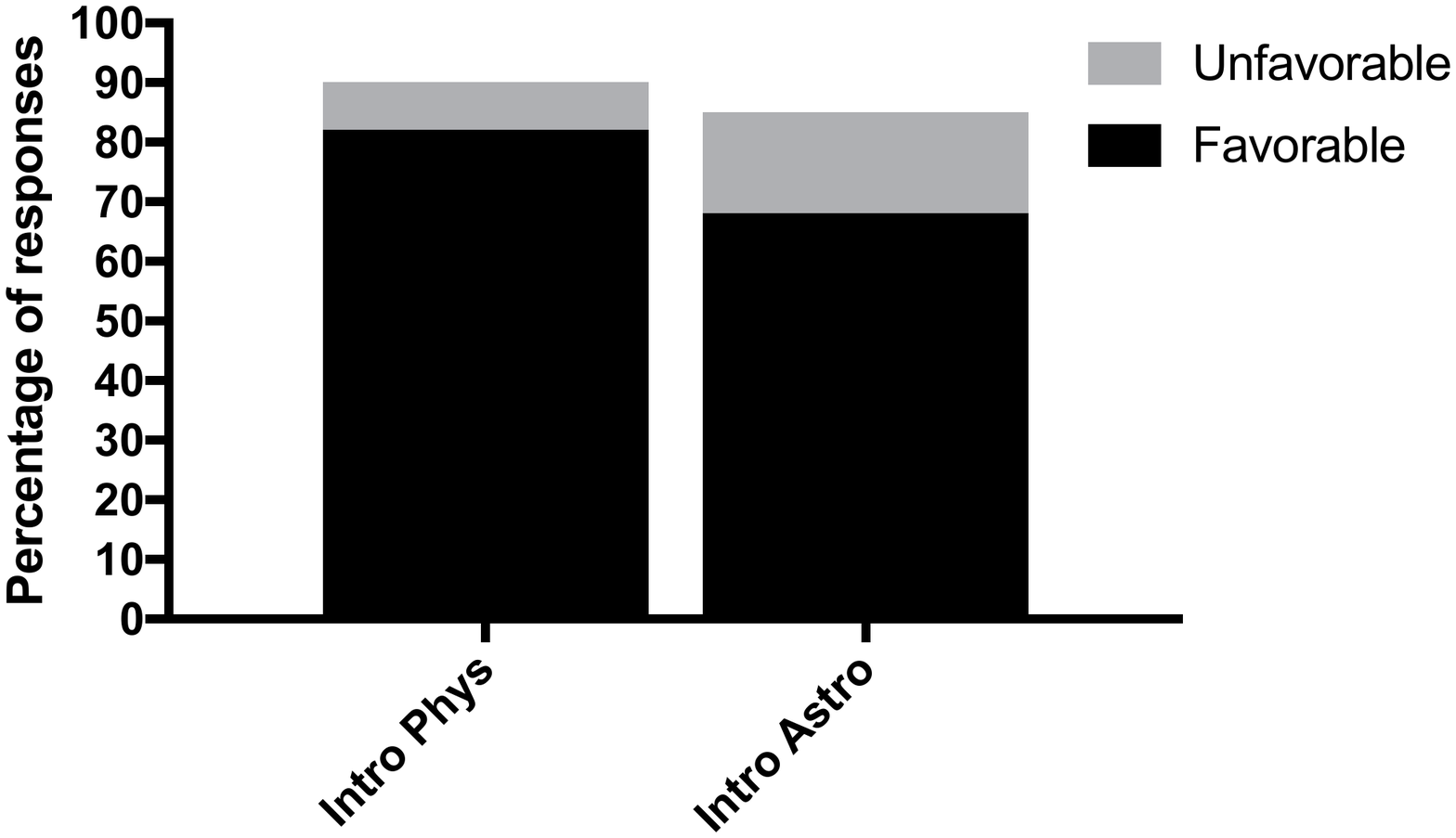}
\caption[justification=raggedright]{Unnormalized scores for Question 15 Astronomy students exhibited \textit{less} favorable attitudes than physics students when asked if they draw pictures or diagrams when solving problems.}
\label{fig:Q15}
\end{figure}

Among the AAPS survey questions on which introductory astronomy students outperformed introductory physics students, the questions with the most significant differences are found in Fig \ref{fig:Q10} - \ref{fig:Q15}, which come from several different factors. Factor 3 involves the perception of problem-solving approach, and astronomy students outperformed physics students in every question except for one question 8 was the exception, in which the average scores were almost identical) within this factor, as seen in Tables \ref{table:Table1}, and \ref{table:astrophyrawQ}. However, neither introductory physics nor introductory astronomy students exhibit a high level of expertise in most of these questions. 

On the other hand, Factor 1, which is related to metacognition and enjoyment in problem solving, suggests higher levels of expertise, and more noticeable differences in scores between introductory physics and astronomy students. This factor was considered to be an important one in the factor analysis in Ref. \cite{Mason1}.  In fact, Factor 1 accounted for more of the variance than any other factor in Ref. \cite{Mason1}. For example, question 10 on the AAPS states ``If I am not sure about the correct approach to solving a problem, I will reflect upon the principles that may apply and see if they yield a reasonable solution.'' Students are supposed to say whether or not they agree with the statements, and a favorable response to this question would be to agree, as 89\% of introductory astronomy students did.  By comparison, only 65\% of introductory physics students gave a favorable response to this question, which can be seen in Figure \ref{fig:Q10}.  As one introductory physics student said in an interview, ``I find it hard, if I was stuck on a problem, to think back on the principles, because if I am struggling then my conceptual understanding isn't that strong.'' Further discussion with this student suggests that this student appeared to believe that it would not be possible to think about the principles if your understanding of them was not strong to begin with.  On the other hand, all but one of the interviewed astronomy students gave a favorable response, stating that they reflect upon principles when they are unsure about the correct approach to solving an astronomy problem, and that the interesting context of these problems is a motivational factor that keeps them engaged even if they are stuck.

Another question from Factor 1, question 27 states, ``I enjoy solving physics/astronomy problems even though it can be challenging at times.'' (Note that physics students only saw the word ``physics'' in this question and astronomy students only saw the word ``astronomy''). As can be seen in Figure \ref{fig:Q27}, while 72\% of astronomy students agreed with this statement, only 40\% of physics students agreed.  One physics student said in an interview, ``To just sit there and do problems over and over is not really fun,'' conveying that she did not find physics to be enjoyable because it involves many problems to solve (that are not intrinsically interesting).  Another student identified the uninteresting nature of physics problems as being why she answered that she did not enjoy physics, stating ``I think it's just like `here's a problem, just solve it' and not really making it an interesting, this could be in the real world thing.'' On the other hand, some interviewed astronomy students explicitly reported that they enjoyed astronomy problems more than physics problems, even if they are just as challenging.  As one student, who had taken both introductory physics and introductory astronomy, put it, ``I think physics and astronomy are strongly aligned for a lot of both of those introductory courses but where they start to split, I just find that subject matter [astronomy] more interesting.  So even though the physics involved can be basically the same, the setup is what drives me to like astronomy more.''  This student indicated that the physics behind the courses is similar, but that the subject matter of astronomy was more enjoyable and kept him engaged and persistent despite the challenging nature of many of the problems.  Other interviewed students expressed similar sentiments.

Related to the metacognitive aspect of Factor 1, some questions asked about conceptual thinking over ``plug-and-chug'' approaches.  For example, question 4 states ``In solving problems in physics/astronomy, I always identify the physics/astronomy principles involved in the problem first before looking for corresponding equations.'' A favorable response to this question would be to agree with the statement, as 80\% of the introductory astronomy student did.  Fewer introductory physics students (62\%) responded favorably by comparison, as seen in Figure \ref{fig:Q4}.  Likewise, question 14 states, ``When I solve physics/astronomy problems, I always explicitly think about the concepts that underlie the problem.''  Figure \ref{fig:Q14} shows that while 78\% of introductory astronomy students indicated a favorable response to this question (i.e., they agreed with the statement), only 55\% of introductory physics students indicated a favorable response.  

Factor 6 was another one in which introductory astronomy students gave more favorable responses than introductory physics students. This factor has to do with problem solving confidence, and relates to students' willingness to persist in working through problems in the face of difficulty.  For example, question 1 says ``If I'm not sure about the right way to start a problem, I'm stuck unless I go see the teacher/TA or someone else for help.'' As can be seen in Figure \ref{fig:Q1}, only 49\% of physics students had favorable responses to this question (i.e., a favorable response would be to disagree with the statement); wheras, 68\% of astronomy students gave favorable responses. Similarly, question 23 states ``If I cannot solve a problem in 10 min, I give up on that problem.''  As shown in Figure \ref{fig:Q23}, a favorable response of disagree was given by 82\% of astronomy students but only 61\% of physics students.

Another factor in which introductory students performed more favorably than introductory physics students was Factor 8, which deals with sensemaking.  For example, question 2 states, ``When solving physics/astronomy problems, I often make approximations about the physical world.''  The majority (65\%) of introductory astronomy students agreed with this statement, constituting a favorable response.  On the other hand, only 43\% of introductory physics students gave a favorable response, as can be seen in Figure \ref{fig:Q2}.  Another question in this factor was question 16, which states, ``When aswering conceptual physics/astronomy questions, I mostly use my `gut' feeling rather than the physics/astronomy principles I usually think about when solving quantitative problems.'' Figure \ref{fig:Q16} shows that while 50\% of introductory physics students gave a favorable response (i.e., they disagreed with the statement), the majority of astronomy students  (71\%) gave a favorable response. 

The only factor in which introductory astronomy students gave \textit{less} favorable responses compared with introductory physics students was Factor 2, which is related to the use of drawings and scratch work while solving problems. For example, question 15 states ``When solving physics/astronomy problems, I often find it useful to first draw a picture or diagram of the situations described in the problems.''  As can be seen in Figure \ref{fig:Q15}, while 82\% of introductory physics students answered favorably (i.e., agreed with the statement) to this question, only 68\% of introductory astronomy students answered favorably. However, in follow-up interviews, one possible reason for less favorable responses was uncovered.  For example, one student described certain topics that come up in astronomy problems to be more difficult to depict in a drawing, stating: ``For example, involving things like a spectrum, or EM [electromagnetic waves] in general.  I’m not going to draw a light wave.''  Another interviewed student touched on a similar theme in making the following distinction about when drawing is helpful in solving astronomy problems, ``For questions about celestial movement, I think that's where the bulk of the drawing comes from...  But like right now  we’re learning about magnitudes, I don't know how I would draw different magnitudes...because I can't draw light...unless it's reflected onto something like the moon.  Phases of the moon I really need to draw.''  This student indicated that there are only some problems in astronomy that appear to lend themselves well to drawing and, similar to other introductory astronomy students who discussed this issue, there are many astronomy problems that were either difficult to depict in a sketched diagrams and/or were problems for which a drawing may not have offered much benefit.  Since introductory physics problems often focus more on simpler physical objects that lend themselves well to drawing, this dichotomy could account for the less favorable responses of astronomy students regarding the use of drawing and scratchwork in problem solving.

\subsection{RQ3.  When presented with an isomorphic problem pair (problems with the same underlying physics principle but different contexts--one written in an astronomy context and the other written in a physics context), are introductory physics and introductory astronomy students equally likely to solve both problems correctly and do they find either more or less interesting?}

At the end of the interviews, out of the 20 total students interviewed, 8 introductory astronomy and 8 physics students were presented with a pair of isomorphic problems (one written in and one without the context of astronomy). Both problems required the use of the concept of centripetal force and the assumption of a uniform circular motion in order to solve for the speed.  Altogether 13 out of 16 of all students (both introductory physics and introductory astronomy) who were presented with these problems were able to solve both problems entirely correctly.  This includes 6 out of 8 of introductory physics students and 7 out of 8 of introductory astronomy students.  Other students in both groups were also able to solve a significant fraction of both isomorphic problems correctly but made some mistakes along the way, e.g., writing centripetal force as $\frac{v^2}{r}$ as opposed to $\frac{mv^2}{r}$.  Like the FCI results, this similarity in problem solving performance on a pair of isomorphic problems suggests that student performance on similar content is comparable for the subset of students interviewed in both classes even though each group has very few students to make statistical inferences. 

While problem solving performance of this set students appears to be similar, differences were found when subjects were asked which of the two isomorphic problems (non-astronomy context or astronomy context) they enjoyed more.  In particular, the majority of both introductory physics and introductory astronomy students reported that they found the astronomy context more interesting.  For example, one physics student, when asked which she found to be more enjoyable reported, ``The Sun and the Earth one, because I think space is cooler than a yo-yo.'' This student indicated that there is something more captivating about space than about everyday objects.  Another physics student stated, ``The Earth one, yes, was more interesting,'' conveying a similar sentiment that the context of the astronomy problem was more interesting than the context of the physics problem. 

It is interesting to note that several students expressed more interest in the astronomy problem, even if they found it to be more difficult to solve or challenging to think about.  For example, one physics student stated, ``[The non-astronomy] one was less annoying to solve because I could think about what the answer could be, but [the astronomy] one, everything's scientific notation so I can't really think about... But the [astronomy] one is definitely more interesting than the first one because it's an application of all these things.'' Similarly, an interviewed astronomy student reported that the physics problem was easier to solve, but that she enjoyed the astronomy problem more.  When asked why the astronomy problem was more enjoyable, she stated, ``I'm just into the subject matter of astronomy.  It's more interesting than a yo-yo's tension,'' again conveying a more motivating scenario and an interest in the astronomy context.

One astronomy student in an interview discussed at length how the astronomy problem sparked more curiosity and interest and described how it got him thinking, ``That's cool to me because then you could think about well what if all of a sudden we stop orbiting and we go off on a tangent, and at that speed, well that sucks, that's going to be rough.''  The context of the astronomy problem appeared to cause him to think further about centripetal force, and how the removal of this force would result in tangential movement at a constant speed. When further asked if the problem with non-astronomy context would inspire him to think about what might happen if the yo-yo string broke and it, similarly, went off on a tangent, he indicated that this thought was not as captivating or interesting, stating that, ``That just means it's hitting my TV and I'm getting yelled at,'' Like this student, most of the interviewed students expressed less interest in the physics problem.

\newpage

\vspace{-0.2in}
\section{Discussion}
\vspace{-0.13in}
\textbf{Similar students, different attitudes:}
 Our results indicate that introductory astronomy students, overall, had more expert-like attitudes and approaches than introductory physics students.  
An explanation for this difference would have to account for the fact that astronomy students had comparable conceptual gain on force concepts as measured by the FCI and a subset of students had comparable performance on two isomorphic problems posed in an astronomy and non-astronomy context although we cannot make statistical inferences about this result (since very few students were asked to solve these isomorphic problems in interviews). The students in the astronomy course consistently gave much more favorable responses to AAPS survey questions involving metacognition and enjoyment in problem solving.  An explanation for the difference between the AAPS scores of introductory physics and introductory astronomy students would also have to account for the fact that both algebra-based and calculus-based introductory physics students score comparably on the AAPS (so much so that their results were combined) \cite{Mason1}.  This is noteworthy because students who enroll in algebra-based physics tend to have different reasons for enrolling in introductory physics than calculus-based physics students, and these two courses contain students which could be argued to be more dissimilar to each other in their prior preparation in mathematics, physics, and abstract thinking, in general, than introductory physics students are to introductory astronomy students in the astronomy course discussed here.  If differences in AAPS score should be expected to be tied to differences in motivation for taking the class, one would expect differences between algebra-based and calculus-based physics courses' average scores to show similar differences, but such a difference was not found.  Different scores at the introductory level appear only when comparing physics students to astronomy students.

\textbf{Captivating content:}
 One possible reason for the differences between the AAPS scores in introductory physics and introductory astronomy may be related to the subject matter in astronomy that might promote higher level of metacognition and may be more interesting and engaging for students compared to that in a typical physics course discussed here. For example, it is possible that learning about galaxies and black holes and viewing images of outer space is more captivating than learning about pulleys and inclined planes. One interviewed student described the appeal of real astronomical images compared with images one might encounter in a physics class as follows, ``For astronomy there are all these real images... you can't take a picture of a projectile the same way, but with astronomy you can take basically any image from the Hubble telescope and be like `there' and that's all you need... What our teacher did was pull from the astronomy picture of the day and it was really cool.  Just having that image really set the tone.'' Even the context of the problems themselves appears to yield more student interest when the problems are written in the context of astronomy, with many students who were interviewed indicating that they enjoyed the astronomy problem more than the isomorphic physics problem. If it is the case that students find astronomy and astronomy problems more interesting, it is possible that more captivating instruction of physics classes, including more engaging contexts such as those used in astronomy classes, may improve physics students' attitudes and approaches to problem solving.  If introductory physics classes can emphasize aspects of physics that resonate with students and capture their interests, their attitudes may more closely resemble those of introductory astronomy students. Using suitable astronomy contexts in physics courses would be one effective approach that physics instructors could consider.

\textbf{Conceptually-rich instruction:}
 Some students identified the conceptually-rich nature of astronomy as a reason they found the instruction interesting.  On the other hand, typical traditional physics course is often taught in a dry manner and problems often emphasize algorithmic problem-solving rather than promoting functional conceptual understanding with engaging content.  An informal survey of introductory physics faculty at the University of Pittsburgh suggests that a majority of the physics homework, quiz, and exam questions instructors assign focus on quantitative problem solving.  Discussions with some instructors suggest that conceptual learning is regarded by physics instructors as ``easier'' or less rigorous than quantitative problem-solving, and that making problems interesting is secondary to developing effective problem-solving skills.  For example, in one qualitative study in which the researcher followed students through the experience of an introductory physics course, it was concluded that when the subject is taught in a way that is ``boring, dull, or simply not fun,'' students will often become disengaged, and that this sort of ``boring'' teaching described the way introductory physics is often taught \cite{Tobias}.  If the content of problems were more engaging through the use of a conceptually rich, interesting context, then perhaps students' attitudes regarding problem-solving would improve. One student who was interviewed hit on this theme when he said ``Algorithmic learning is kind of terrible,'' as he described the way in which physics is taught in his opinion.  He went on to contrast this with the way astronomy was taught, saying ``...conceptual learning really builds foundation for the subject.'' This student then identified ways in which physics problem-solving could be improved,  ``if there were pointed questions that went along with the mathematics, like even a mathematics question and then at the very end ``why?''  He felt that physics courses only focused on ``plug and chug'' and there was a de-emphasis on integrating the conceptual aspects of the problem in the student learning process.

Another interviewed student expressed that physics courses could engage students more in problem-solving through better use of demonstrations to build conceptual thinking.  As this student explained, ``I think in physics there's more possibility for demonstrations... The one thing though was that the demos...some of them were really cool like when we were learning about angular momentum and you're spinning in a chair, but some of them tended to be so technical that you lost interest.  But I think if physics took a lot more advantage of the demos, the ability to do those kind of demos, if students could really see what is going on, in a really spectacular way, that would be like `oh yeah!'''   This sort of active engagement has been shown to be a benefit of interactive lecture demonstrations \cite{Sokoloff}

After discussing the idea of combining conceptual instruction with problem-solving, one student explained that physics instruction could be made more interesting by including problems that would engage students in something fun, after conceptual instruction, stating: ``Here's an idea.  You take famous scenes from movies and you try to figure out what's actually happening in them.  So if you have a James Bond car chase scene and you see the car spinning in a circle and you're teaching friction forces, you could say if the car spins out 10 meters that way and assuming there's no outside forces, what is the friction force on the tires?'' This student saw the power in bringing concepts to life by using exciting problems. Indeed, combining conceptual reasoning with problem-solving may help promote problem-solving expertise, \cite{Mazur,Lin2} 
which could translate to more favorable attitudes and approaches to problem solving. Therefore, incorporating more conceptually-rich instruction into the physics classroom could be an important piece of the puzzle. 

\textbf{Impact of motivation on learning:} 
Motivational characteristics of students is often divided into a few broad categories, e.g., motivational characteristics of those who are focused mainly on performing well in a class without much focus on achieving understanding, i.e., a ``performance'' motivation orientation, or those who are focused on achieving  
mastery of the material, i.e., a ``mastery'' motivation orientation \cite{Belenky, Pugh,Nokes}. These different motivational goals can have a negative or positive feedback loop effect on student attitudes and may shape their problem-solving beliefs and processes through different levels of engagement and sense-making while solving problems\cite{Nokes}.

While the student body is similar in physics and astronomy courses in this investigation, many students take the astronomy course as an elective. This may point to an intrinsic motivation to take this course as opposed to an extrinsic motivation for the physics course, which is required for most students. We note however that while students taking a class as an elective or as a requirement could  be expected to be one factor in their motivational characteristics at the end of the course, we find that it is possible to acquire similar conceptual learning gain, e.g., in force concepts (as measured by average FCI pre-/post-scores), for both groups.

Because many AAPS survey questions tap into issues of motivational goals, 
one  
hypothesis inspired by our finding is that the more favorable responses in introductory astronomy at the end of the course may imply that not only are students more motivated to learn the astronomy content from the beginning but the content encourages a more mastery-oriented motivational goal and keeps students in a positive feedback loop so that their attitudes remain higher than those in introductory physics at the end of the course. 
This hypothesis will be tested in a future investigation in which we will administer the AAPS survey both as a pre and posttest to both sets of students.
In particular, these findings suggest the possibility that instruction of 
the mandatory introductory physics classes could potentially be tailored to more positively motivate students without sacrificing rigor in learning. Since there is a significant overlap in the type of students taking introductory physics and  astronomy, and since their gain in conceptual measures on force concepts measured by the FCI is equivalent, it will be useful to uncover whether the discrepancy in their AAPS scores could at least partly be addressed through the ways in which introductory physics is traditionally taught. 
 In particular, if introductory physics students can be motivated to learn and more effectively engaged with the content, as it appears introductory astronomy students are, then attitudes and approaches to problem solving for introductory physics students may improve \cite{MarshmanMot}.

\section{Summary and Implications}

Written survey responses and interviews suggest that at the end of the course, introductory astronomy students had more favorable attitudes and approaches to problem solving overall, and in the majority of clusters of individual questions. In light of the results presented here pertaining to the AAPS survey scores in introductory physics and introductory astronomy courses and individual interviews with students, an important instructional implication of the more favorable attitudes and approaches to problem-solving found among introductory astronomy students compared to introductory physics students is for instructors of physics to consider incorporating conceptually-rich and engaging content in their courses.  For example, the problems that physics students are asked to solve could include more exciting and realistic contexts, or involve objects that they can relate to or find interesting.  Preceding problem-solving with interactive demonstrations to build conceptually-rich instructional design in which quantitative and conceptual aspects of learning are integrated may also motivate students to relate to subject matter involved in the problems they are asked to solve. Additionally, introductory physics instructors could utilize the fascination students often have with the cosmos in their teaching of introductory physics by incorporating astronomical objects and examples into problems when appropriate. Moreover, problems which contain real images rather than simply cartoons or sketches may bring the physics content to life for students. From in-class examples, to homework problems, to exam questions, if students can more effectively engage with the content and questions asked of them, as it appears astronomy students often do, then  the attitudes and approaches to problem solving for introductory physics students has the potential to improve.

\begin{acknowledgments}

We thank the NSF for award DUE-1524575.
\end{acknowledgments}

\newpage

\begin{table}
\caption{\small Effect sizes between different groups}
\label{table:Table0}
\centering
\fontsize{10}{12}\selectfont
\begin{tabular}{|m{5em}| c | c | c |}
\hline 
\textbf{Cohen's d}  & \textbf{Physics} & \textbf{Graduate} & \textbf{Faculty}
\\
\hline
Astronomy & 0.81 & 1.06 & 1.78 
\\
Physics & & 1.60 & 2.18 
\\
Graduate & & & 1.19
\\
Faculty & & & 
\\
\hline
\end{tabular}
\end{table}

\begin{table}[!htbp]
\caption{\small Average normalized scores by question}
\label{table:Table1}
\centering
\fontsize{10}{12}\selectfont
\begin{tabular}{|m{10em}| c | c | c | c |c | c | c |}
\hline 
\textbf{Question number}  & \textbf{1} & \textbf{2} & \textbf{3} & \textbf{4} & \textbf{5} & \textbf{6} & \textbf{7}
\\
\hline
Introductory physics & 0.14 & 0.19 & 0.15 & 0.41 & 0.16 & 0.24 & 0.61
\\
Introductory astronomy & 0.40 & 0.49 & -0.15 & 0.71 & 0.23 & 0.51 & 0.83
\\
Graduate students & 0.71 & 0.42 & -0.04 & 0.83 & 0.17 & 0.75 & 0.83
\\
Faculty & 0.83 & 1.00 & 0.50 & 0.92 & 0.92 & 1.00 & 0.83
\\
\hline
\textbf{Question number} & \textbf{8} & \textbf{9} & \textbf{10} & \textbf{11} & \textbf{12} & \textbf{13} & \textbf{14}
\\
\hline
Introductory physics & 0.67 & 0.24 & 0.58 & -0.03 & -0.06 & 0.56 & 0.32
\\
Introductory astronomy & 0.66 & 0.31 & 0.88 & 0.26 & 0.13 & 0.75 & 0.72 
\\
Graduate students & 0.83 & 0.46 & 0.88 & 0.67 & 0.54 & 0.88 & 0.88 
\\
Faculty & 0.92 & 0.58 & 0.92 & 0.67 & 0.83 & 1.00 & 0.50
\\
\hline
\textbf{Question number} & \textbf{15} & \textbf{16} & \textbf{17} & \textbf{18} & \textbf{19} & \textbf{20} & \textbf{21}
\\
\hline
Introductory physics & 0.74 & 0.23 & 0.55 & 0.69 & 0.77 & -0.19 & 0.71
\\
Introductory astronomy & 0.51 & 0.58 & 0.28 & 0.48 & 0.86 & 0.26 & 0.86
\\
Graduate students & 0.96 & 0.50 & 0.79 & 0.96 & 0.88 & 0.38 & 0.92
\\
Faculty & 1.00 & 0.67 & 1.00 & 0.92 & 1.00 & 0.75 & 1.00
\\
\hline
\textbf{Question number} & \textbf{22} & \textbf{23} & \textbf{24} & \textbf{25} & \textbf{26} & \textbf{27} & \textbf{28}
\\
\hline
Introductory physics & 0.52 & 0.40 & 0.43 & 0.56 & 0.37 & 0.03 & 0.75
\\
Introductory astronomy & 0.69 & 0.75 & 0.15 & 0.71 & 0.60 & 0.65 & 0.89
\\
Graduate students & 1.00 & 1.00 & 0.21 & 0.54 & 0.71 & 0.67 & 0.96
\\
Faculty & 1.00 & 0.92 & 0.42 & 0.92 & 1.00 & 0.92 & 1.00
\\
\hline
\textbf{Question number} & \textbf{29} & \textbf{30} & \textbf{31} & \textbf{32} & \textbf{33} & \textbf{Overall} & 
\\
\hline
Introductory physics & 0.74 & -0.04 & 0.08 & 0.70 & 0.46 & 0.38 & 
\\
Introductory astronomy & 0.82 & 0.14 & 0.12 & 0.80 & 0.63 & 0.53 & 
\\
Graduate students & 1.00 & 0.92 & 0.92 & 1.00 & 0.83 & 0.72 &
\\
Faculty & 1.00 & 1.00 & 1.00 & 1.00 & 1.00 & 0.88 &
\\
\hline
\end{tabular}
\end{table}

\begin{table}[t]
\includegraphics[height=0.95\textheight]{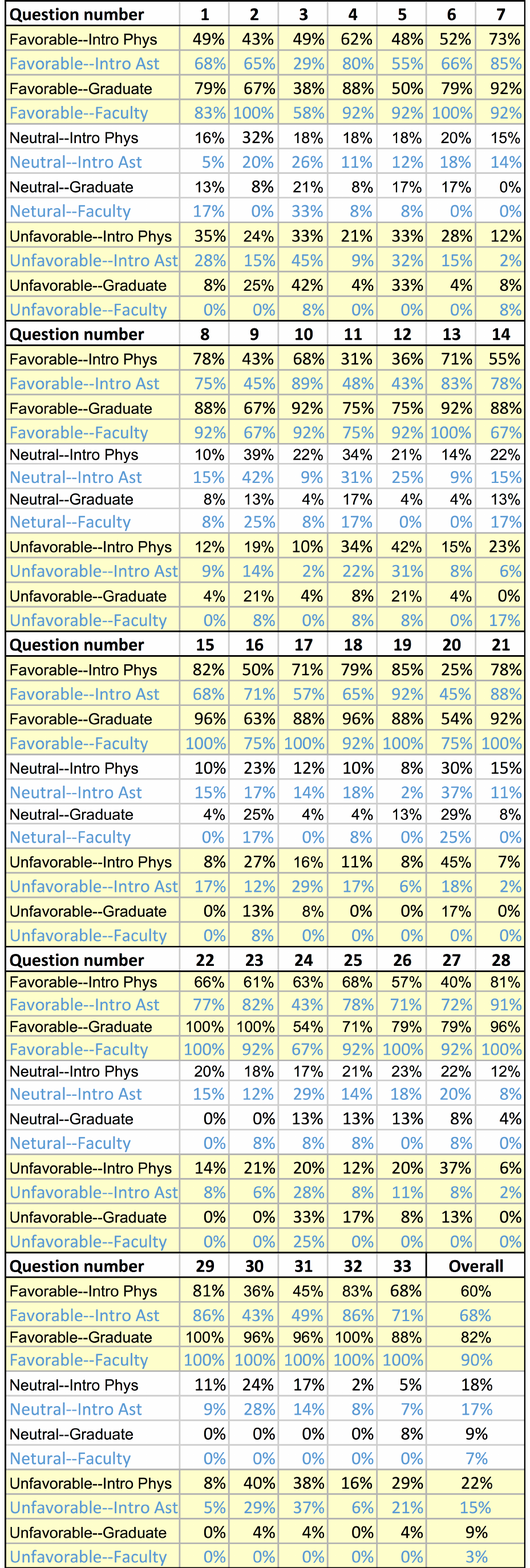}
\caption[justification=raggedright]{Average unnormalized scores for individual questions showing the breakdown of average favorable, unfavorable, and neutral responses to individual questions of the AAPS.}
\label{table:astrophyrawQ}
\end{table}

\newpage

\begin{table}
\caption{\small Average normalized scores by question and factor.  Order of question numbers reflects that in Ref. \cite{Mason1}.}
\label{table:Table2}
\includegraphics[width=0.5\textwidth]{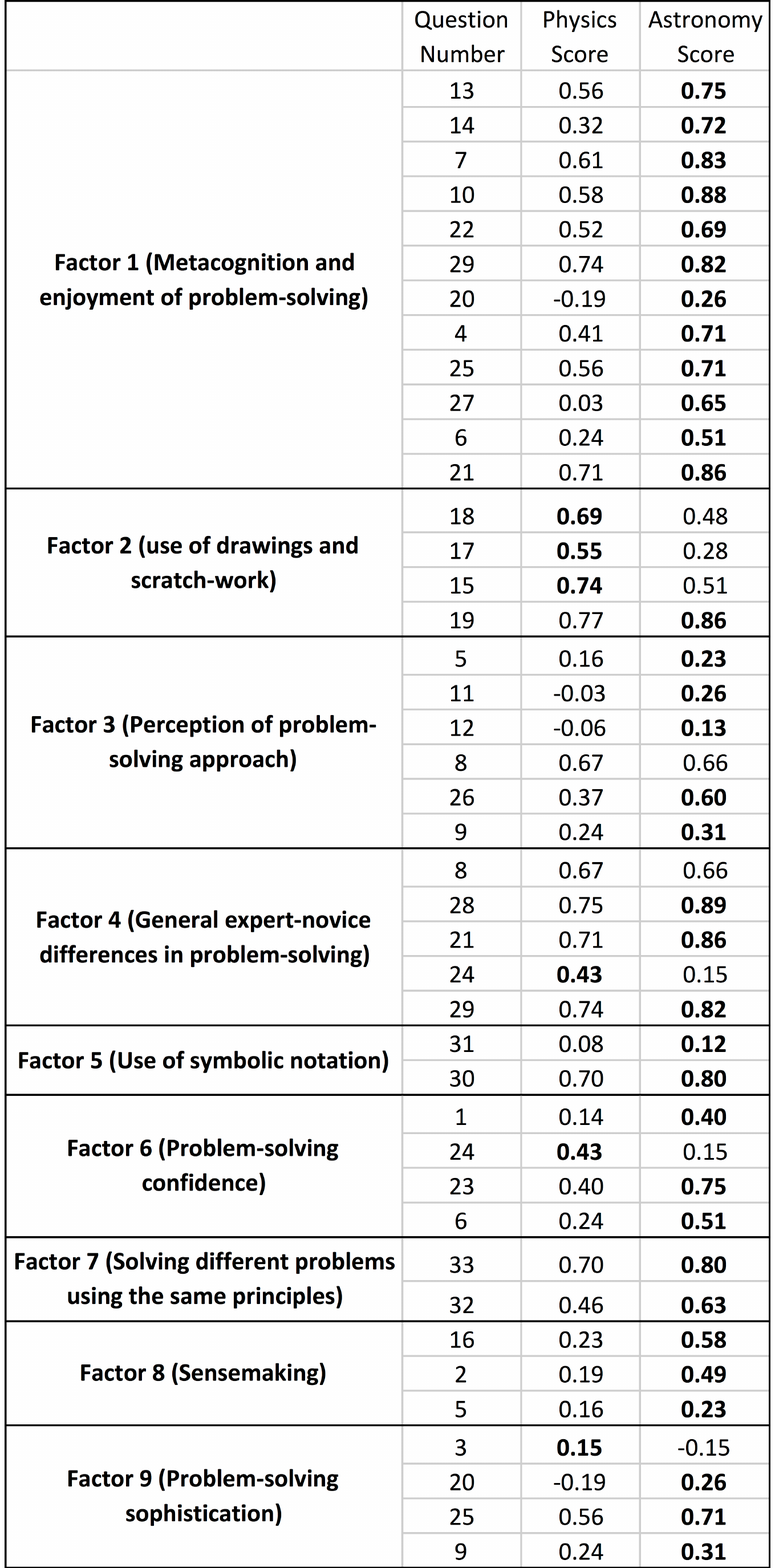}
\end{table}

\end{document}